\documentclass[aps,pra,reprint,floatfix]{revtex4-1}

\pdfoutput=1
\usepackage[latin1]{inputenc}
\usepackage[english]{babel}
\usepackage{latexsym}
\usepackage{graphicx}
\usepackage{amsmath}
\usepackage{natbib}

\newcommand{\rp}{$\langle r_p\rangle$}

\newcommand{\Hp}{e$^+$H}
\newcommand{\Hep}{e$^+$He}
\newcommand{\Lip}{\lowercase{e}$^+$\uppercase{L}\lowercase{i}}
\newcommand{\Bep}{\lowercase{e}$^+$\uppercase{B}\lowercase{e}}
\newcommand{\emthree}{$\times$10$^{-3}$}
\newcommand{\Veff}{$V_{eff}$}

\newcommand{\VeffPsA}{$V_{eff'}^{Ps}$}
\newcommand{\Eeff}{$E_{eff}$}
\newcommand{\Meff}{$M_{eff}$}
\newcommand{\me}{$m_e$}
\newcommand{\mPs}{$2m_e$}

\newcommand{\VBN}{$V^{BN}_{ep}$}
\newcommand{\EepTwoC}{$E_{ep}\left[n_+,n_-\right]$}
\newcommand{\VepTwoC}{$V_{ep}\left[n_+,n_-\right]$}

\newcommand{\PhaseShift}{$\delta_0$}
\newcommand{\ScatLength}{$A_0$}

\begin{document}

\title{
Full correlation single-particle positron potentials for a positron and a positronium interacting with atoms
}
\author{A. Zubiaga}
\email{asier.zubiaga@gmail.com}
\author{F. Tuomisto}
\affiliation{Department of Applied Physics, Aalto University, P.O. Box 14100, FIN-00076 Aalto Espoo, Finland}
\author{M.~J. Puska}
\affiliation{COMP, Department of Applied Physics, Aalto University, P.O. Box 11100, FIN-00076 Aalto Espoo, Finland}

\begin{abstract}
In this work we define single-particle potentials for a positron and a positronium atom interacting with light atoms (H, He, Li and Be) by inverting a single-particle Schr\"odinger equation. For this purpose, we use accurate energies and positron densities obtained from the many-body wavefunction of the corresponding positronic systems. The introduced potentials describe the exact correlations for the calculated systems including the formation of a positronium atom. We show that the scattering lengths and the low-energy s-wave phase shifts from accurate many-body calculations are well accounted for by the introduced potential. We also calculate self-consistent two-component density-functional theory positron potentials and densities for the bound positronic systems within the local density approximation. They are in a very good agreement with the many-body results, provided that the finite-positron-density electron-positron correlation potential is used, and they can also describe systems comprising a positronium atom. We argue that the introduced single-particle positron potentials defined for single molecules are transferable to the condensed phase when the inter-molecular interactions are weak. When this condition is fulfilled, the total positron potential can be constructed in a good approximation as the superposition of the molecular potentials. 
\end{abstract}

\keywords{positronium chemistry, stochastic variational method, effective potential}

\maketitle

\section{Introduction}
Although the chemistry of the positron in crystalline solids and soft-condensed matter has an intrinsic interest by itself, it is mainly studied in connection of probing the electron chemistry and the open volume of materials by positrons. Thermalized positrons become localized inside open volume defects such as vacancies and voids where the repulsion by the nucleus is minimum. When probing soft matter, the positron chemistry has to be taken into account because a positron can bind an electron and form a positronium (Ps) atom before getting trapped into open volume pockets~\cite{Book_Mogensen}. The annihilation properties of the positron are determined by the local electronic structures and the distribution of the open volume. Positron annihilation spectroscopy (PAS) exploits this property to measure the type and concentration of vacancies in metals and semiconductors~\cite{RMP_Tuomisto}. By measuring the lifetime of Ps, the distribution of open volume has been studied in porous SiO$\mathrm{_2}$~\cite{PRB_Nagai, APL_Liszkay}, polymers~\cite{JPS_Uedono} and biostructures~\cite{JPCBL_Sane}. 

The interpretation of PAS experiments benefits from the comparison to computational predictions. However, the description of an electron-positron system embedded in a host material requires addressing the correlations of light particles beyond the adiabatic approximation, the quantum-mechanical delocalization and the zero-point energy. Regrettably, using many-body techniques for a full quantum-mechanical treatment of the problem is clearly beyond the present-day computational capacity. Instead, in metals and semiconductors the distributions and the annihilation properties of positrons can be calculated from first principles to a good accuracy within the two-component density functional theory (2C-DFT) and the local density approximation (LDA) for the exchange and correlation functionals~\cite{RMP_Puska}. For a delocalized positron in a perfect lattice the scheme works particularly well because the electron-positron correlation energy functional is known very accurately within the LDA in this limit. Moreover, the same method can be applied also for positrons trapped at vacancies. The calculated positron annihilation parameters can then be used for a quantitative analysis of the experimental results for metals and semiconductors~\cite{RMP_Tuomisto}. However, the 2C-DFT scheme is considered to be unable to describe Ps and instead semiempirical methods have been employed to describe the matter-Ps interaction~\cite{JCP_Tao, CP_Eldrup, JCP_Schmitz}.

The many-body wavefunctions of small positronic systems composed by a positron interacting with a light atom or a small molecule can be calculated to a good accuracy using the Quantum Monte-Carlo (QMC)~\cite{JCP_Kita1,JCP_Kita2, JCP_Bressanini1, JCP_Bressanini2, JCP_Bressanini3, JCP_Mella1} and Configuration Interaction (CI)~\cite{JCP_Buenker} methods. In this work, we have obtained accurate positron energies and densities for positronic atoms including a positron (\Hp{}, \Hep{}, \Lip{} and \Bep{}) and Ps (HPs and LiPs) by an exact diagonalization stochastic variational method (SVM) using an explicitly correlated Gaussian (ECG) function basis set. 

In \Lip{} an electron from Li forms a Ps atom with the positron and becomes bound to the Li$^+$ ion. In \Bep{} the polarized electron cloud binds the positron. In HPs and LiPs the unpaired atom electrons form a chemical bond that binds the Ps strongly to the atom. On the other hand, the positron is not bound to the atom in the \Hp{} and \Hep{} systems. ECG-SVM accounts for more correlation energy for the bound states~\cite{PRA_Mitroy5,JAMS_Mitroy,PRA_Bubin} and the resulting binding energies are larger than for QMC and CI. 

On the basis of our many-body results we propose a single-particle potential for the positron and we derive it for all the positronic systems by inverting a single-particle Schr\"odinger equation. We check the accuracy by comparing the ensuing scattering lengths and the s-wave phase shifts to the corresponding many-body values.  We also compare the many-body densities and the introduced single-particle potentials to the corresponding results of 2C-DFT within LDA for bound \Lip{}, \Bep{}, HPs and LiPs. The agreement seen predicts that 2C-DFT and LDA can be the starting point to describe positron bound states including systems in which Ps is formed. Finally, we discuss the utility of the single-particle effective potentials to approach a practical and predictive description of positron and Ps states in condensed matter. 

The organization of the present paper is as follows. The many-body ECG-SVM as well as the 2C-DFT schemes are shortly described in Chapter~\ref{sec_computational}. Chapter~\ref{sec_results} presents and discusses the effective single-particle potentials, the elastic scattering parameters and the self-consistent 2C-DFT-LDA results. Chapter~\ref{sec_discussion} is devoted to discuss the utility of the introduced potentials to describe positron and Ps states in condensed matter and chapter~\ref{sec_conclusions} presents our conclusions. 

\section{Computational methods}~\label{sec_computational} 
\subsection{ECG-SVM}
ECG-SVM~\cite{PRC_Varga} is an all-particle quantum ab-initio method used to calculate the many-body wavefunction of N particles (electrons, positrons and nuclei) interacting through the Coulomb interaction. The Hamiltonian of the system with the kinetic energy of the center-of-mass (CM) subtracted is
\begin{equation}
H = \sum_i\frac{p_i^2}{2m_i} - T_{CM} + \sum_{i<j}\frac{q_iq_j}{4\pi\epsilon_0r_{ij}},
\end{equation}
where $\vec{p}_i$ is the momentum, $m_i$ the mass, and $q_i$  the charge of the i$^{th}$ particle, $r_{ij}$ is the distance between the $i^{\rm th}$ and $j^{\rm th}$ particles and $T_{CM}$ the kinetic energy of the CM. The hadronic nucleus is treated as a point particle without structure, on equal footing with the electrons and the positron. The wavefunction is expanded in terms of a linear combination of properly antisymmetrized ECG functions, 
\begin{eqnarray}\label{eq1}
\Psi(x) 
= \sum_{i=1}^s c_i\ 
{\displaystyle \mathcal{A} } 
\left [ 
\exp
^{-\frac{1}{2}xA^ix}
\right ]
\otimes\chi_{SMs} 
,
\end{eqnarray}
where $A^i$ are the non-linear coefficient matrices and $c_i$ the mixing coefficients of the eigenvectors. The antisymmetrization operator $\mathcal{A}$ acts on the indistinguishable particles and $\chi_{SMs}$ is a spin eigenfunction with $\hat{S}^2\chi_{SMs}=S(S+1)\hbar^2\chi_{SMs}$ and $\hat{S}_z\chi_{SMs}=M_S\hbar\chi_{SMs}$. The ECG basis uses Jacobi coordinate sets \{$x_1$,...,$x_{N-1}$\} with the reduced mass $\mu_i$ = $m_{i+1}\sum_{j=1}^im_j/\sum_{j=1}^{i+1} m_j$ that allows for a straightforward separation of the CM movement. All the systems we have considered so far have zero total angular momentum, so we do not need to include spherical harmonics to describe the orbital motion. The electron density is $n_-(r) = \sum_{i=1}^{N_e}\langle\Psi |\delta(\vec{r}_i-\vec{r}_N-\vec{r})|\Psi\rangle$ and the positron density is $n_{+}(r) = \langle\Psi |\delta(\vec{r}_p-\vec{r}_N-\vec{r})|\Psi\rangle$, where $\vec{r}_i$, $\vec{r}_p$, and $\vec{r}_N$ are the coordinates of the i$^{th}$ electron, the positron and the nucleus, respectively.  

The ECG basis sets used in this work comprise between 200 and 2000 functions. Typically, systems with more particles need larger function basis sets for an accurate determination of the wavefunction. The non-linear coefficients $A^i_{\mu\nu}$ are to be optimized to avoid very large basis sets. SVM, which is better suited for functions with a large number of parameters than direct search methods, is used for this purpose. The values of the parameters are varied randomly and the new values are kept only if the update lowers the total energy of the system. The success of the ECG-SVM method relies on the efficient calculation of the matrix elements. 

\lowercase{e}$^+$\uppercase{B}\lowercase{e} converges noticeably slower than other systems of similar size (\lowercase{e}$^+$\uppercase{L}\lowercase{i} and LiPs). . We obtain 2.33\emthree{}~a.u. for the positron binding energy in \Bep{}, while the most accurate value from the literature is 3.163\emthree{}~a.u.~\cite{JAMS_Mitroy}. On the other hand, the dissociation energy of \Lip{} against Li$^+$ and Ps, 2.42\emthree{}~a.u., and the Ps binding energy of LiPs, 11.011\emthree{}~a.u., are closer to the ECG-SVM reference values, 2.4821\emthree~a.u.~\cite{PRA_Mitroy5} and 12.371\emthree{}~a.u.~\cite{JAMS_Mitroy}, respectively. Finally, we obtain 39.187\emthree~a.u. for the Ps binding energy in HPs, in excellent agreement with the reference value of 39.19\emthree{}~a.u.~\cite{PRA_Bubin}. 
\begin{table*}[t!]
\caption{
Main properties of the calculated systems. The first four columns give the name of the system, the size of the basis used, the total energy, and the mean positron-nucleus distance \rp{}. The next three columns give the asymptotic state, the total energy of the corresponding atom or ion and the interaction energy. 
}
\label{tab1}
\begin{ruledtabular}
\begin{center}
\begin{tabular}{cccccccc}
System&Basis size&Energy&\rp{}&Asymptotic&Atom/Ion&E$_{int}^{e+}$/E$_{int}^{Ps}$\\
&&(a.u.)&(a.u.)&state&Energy (a.u.)&(a.u.)\\
\hline
\Hp{}&200&-0.49974&67.47&e$^+$&-0.5 (H)&0.262$\times$10$^{-3}$\\
\Hep{}&1000&-2.90332&56.98&e$^+$&-2.9036937 (He)&0.372$\times$10$^{-3}$\\
\Bep{}&2000&-14.6694&10.972&e$^+$&-14.6670283~(Be)&-2.33$\times$10$^{-3}$\\
&&&&&-14.3246131~(Be$^+$)&\\
\Lip{}&1000&-7.53226&9.928&Ps&-7.2798377~(Li$^+$)&-2.42$\times$10$^{-3}$\\
HPs&1000&-0.78919&3.662&Ps&-0.25 (Ps)&-39.187$\times$10$^{-3}$\\
LiPs&2000&-7.73898&6.432&Ps&-7.4779733~(Li)&-11.011$\times$10$^{-3}$
\end{tabular}
\end{center}
\end{ruledtabular}
\end{table*}

Unbound \Hp{} and \Hep{} can be calculated variationally adding an external confining potential. We have used a weak two-body attractive potential,  
\begin{equation}
V(r_p) = \left\{
\begin{array}{l@{\ \ ,\ \ \ }r}
0&r<R_0\\
\alpha(r_p-R_0)^2&r\ge R_0,
\end{array}
\right.
\end{equation}
binding the positron to the hadronic nucleus in a similar fashion to the confinement potential used by Mitroy et al. to describe positrons scattering off atoms~\cite{PRL_Mitroy}. The potential is different from zero only when the nucleus-positron distance $r_p$ grows above the boundary value $R_0$, and then it has a parabolic increase. $R_0$ and $\alpha$ were set so that the average nucleus positron distance \rp{} $\gtrsim$ 50~a.u.. The confinement radius is chosen large enough (100~a.u.) so that the shape of the wavefunction is not affected by the confinement potential in the interaction region of the positron or Ps with the atom. The resulting \rp{} is large and the interaction energy is small. 

We defined the asymptotic non-interacting state of the atom-positron and atom-Ps systems when the positron is far from the atom. For unbound systems, the asymptotic state is the main scattering channel, i.e., the positron scatters off the neutral atom, and for bound systems it is the main dissociation channel, i.e., the positron (\Bep{}) or Ps (\Lip{}, HPs and LiPs) detaches leaving behind an atom or an ion. \Bep{} splits into a neutral atom and a positron, \Lip{} splits into a Li$^+$ ion and a Ps atom and both HPs and LiPs dissociate into neutral atoms and Ps. We define the positron interaction energy as $E^{e+}_{int} = E_{e^+X} - E_X$, i.e., the difference between the energy of the interacting positronic system, $E_{e^+X}$, and the atom without the positron, $E_X$. The Ps interaction energy, $E^{Ps}_{int} = E_{e^+X/XPs} - E_{X^+/X} - E_{Ps}$ is the difference between the energy of the interacting system, $E_{e^+X/XPs}$, and the sum of the total energies of the positive ion or the atom, $E_{X^+/X}$, and Ps, $E_{Ps}$, after Ps has dissociated.

\subsection{Two-component DFT}
\begin{figure*}
\begin{center}
\includegraphics[width=16cm]{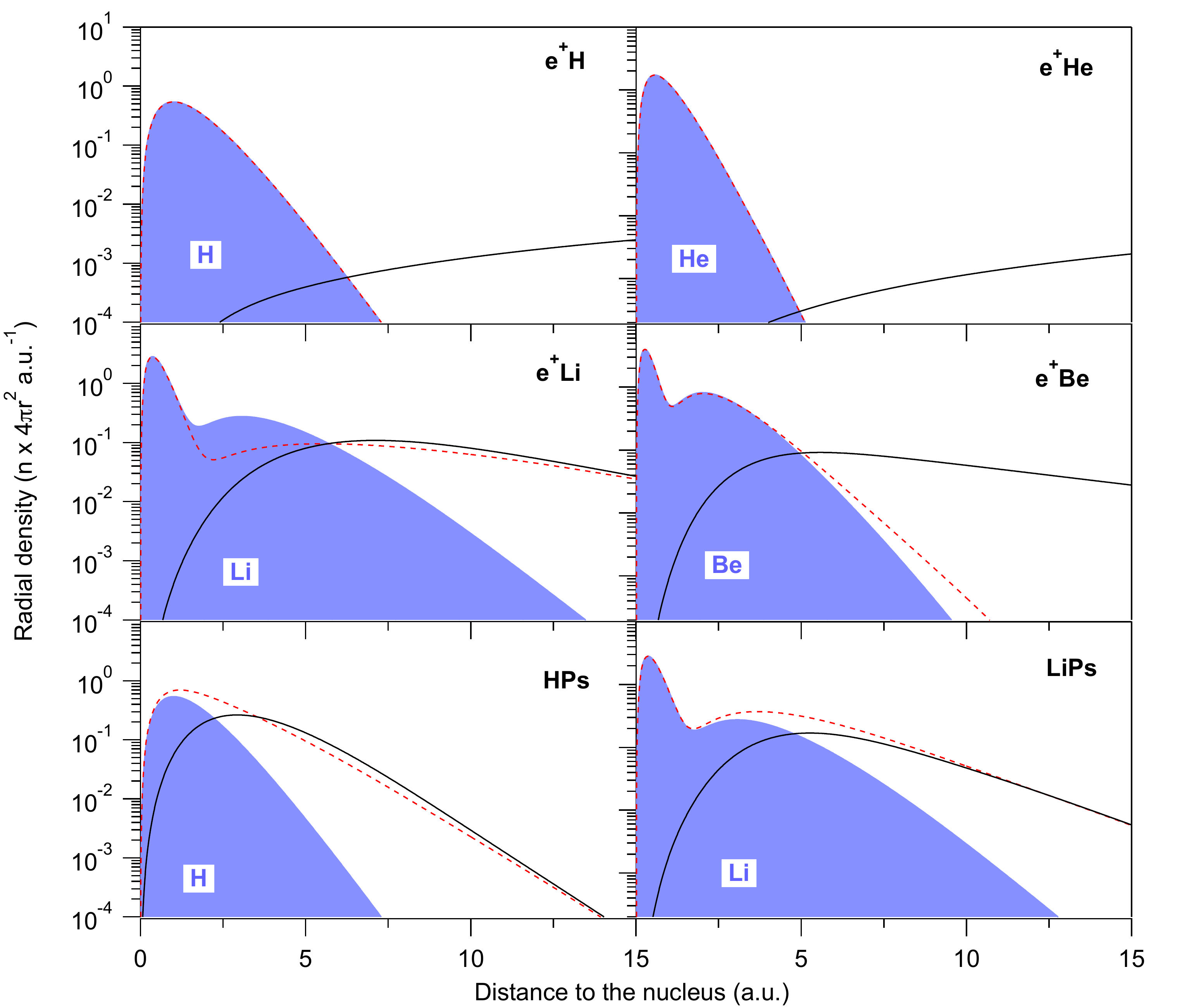}
\caption{(Color online) Electron and positron densities of the calculated systems. The electron densities of the isolated atoms (filled blue curves), and the interacting positron-atom systems (red broken curves), as well as the positron densities (black full curves) are given.} 
\label{fig1}
\end{center}
\end{figure*}
\begin{figure*}
\begin{center}
\includegraphics[width=16cm]{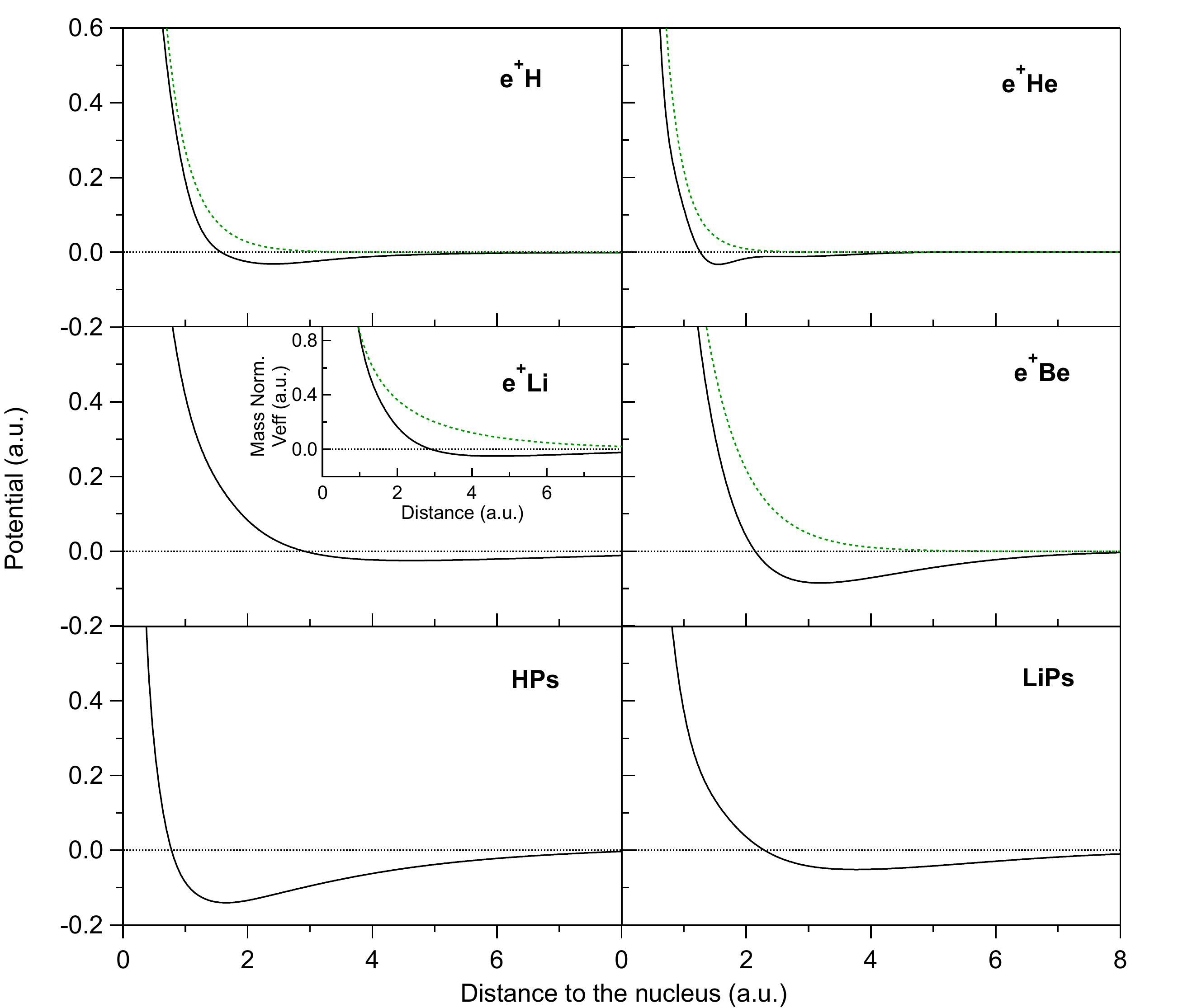}
\caption{(Color online) \Veff{} (black curves) of all the calculated systems and the mean-field Coulomb potentials (green dotted curves) of the systems composed by an atom and a positron. For \Lip{}, HPs and LiPs the Ps \Veff{} have been plotted. The inset in the \Lip{} panel compares the mass normalized \Veff{} and the mean-field Coulomb potentials. } 
\label{fig3}
\end{center}
\end{figure*}

Within LDA of the 2C-DFT the total energy functional of a positronic atom is
\begin{eqnarray}
&&E\left[n_-(r), n_+(r)\right]= \nonumber\\
&&F_1\left[n_-\right] + F_2\left[n_+\right]
+E^{ep}_{C}\left[n_-,n_+\right] + E^{ep}_{corr}\left[n_-,n_+\right],
\end{eqnarray}
where $E^{ep}_{C}\left[n_-,n_+\right]$ is the attractive mean-field Coulomb interaction between the electrons and the positron and $E^{ep}_{corr}\left[n_-,n_+\right]$ is the electron-positron correlation energy functional. $F_1\left[n\right]$ is the usual one-component density functional
\begin{equation}
F_1\left[n\right] = E_{kin}\left[n\right] + E_{ext}\left[n\right] + E_H\left[n\right] + E_{xc}\left[n\right], 
\end{equation}
where $E_{kin}\left[n\right]$ is the Kohn-Sham kinetic energy and $E_{ext}\left[n\right]$, $E_H\left[n\right]$, and $E_{xc}\left[n\right]$ are the electron(positron)-nucleus interaction, the Hartree energy functional and the exchange-correlation energy functional, respectively. For the last one, we have used the parametrization by Perdew and Zunger~\cite{PRB_Perdew}. The self-interaction corrected (SIC) density functional for a single positron $F_2\left[n\right]$ is 
\begin{equation}
F_2\left[n\right] = E_{kin}\left[n\right] + E_{ext}\left[n\right]. 
\end{equation}
The asymmetric treatment of the electron and positron self-interactions for positron states in solids has been shown to give results in a quantitative agreement with experiments~\cite{PRB_Boronski, PRB_Puska3}. The resulting Kohn-Sham equations for the electron $\phi_i^-$ and positron $\phi^+$ orbitals are
\begin{equation}\label{eq_2ddft_1}
\begin{split}
\left[ -\frac{\nabla^2}{2} \right. & - \frac{Z}{r} + \int  \frac{n_-(x) - n_+(x)}{|\vec{r}-\vec{x}|}d\vec{x} 
\\
&\left. + \frac{\delta E_{xc}[n_-]}{\delta n_-} + \frac{\delta E^{ep}_{corr}[n_+,n_-]}{\delta n_-} \right]\phi^-_i = \epsilon^-_i \phi^-_i
\end{split} 
\end{equation} 

\begin{equation}\label{eq_2ddft_2}
\begin{split}
\left[-\frac{\nabla^2}{2} + \frac{Z}{r} \right.&-\int \frac{n_-(x)}{|\vec{r}-\vec{x}|}d\vec{x} 
\\
&\qquad {} \left.
+ \frac{\delta E^{ep}_{corr}[n_+,n_-]}{\delta n_+} \right]\phi^+ = \epsilon^+\phi^+,
\end{split} 
\end{equation} 
where Z is the atomic number of the nucleus and $\epsilon_i^-$ and  $\epsilon^+$ are the electron and positron energy eigenvalues, respectively. Equations~\ref{eq_2ddft_1} and~\ref{eq_2ddft_2} are solved self-consistently with a DFT code that solves the all-electron and positron radial Kohn-Sham equations~\cite{QE-2009}. The mean-field Coulomb potential plotted in figure~\ref{fig3} is composed by the second and third terms of equation~\ref{eq_2ddft_2}. 

In our LDA energy functional we use a two-component electron-positron correlation energy functional \EepTwoC{}. To build up this functional, there is only data for a homogeneous electron-positron plasma in the metallic regime calculated by Lantto~\cite{PRB_Lantto}. The LDA parametrization of reference~[\onlinecite{PRB_Puska3}] describes correctly \EepTwoC{} and \VepTwoC{}=$\delta$\EepTwoC{}/$\delta n_+$ for the electron densities typical in metals and semiconductors $r_s = (3/4/\pi/n)^{1/3} \sim$ 3~a.u. \EepTwoC{} is not known accurately at medium electron and positron densities, beyond $r_s$~$>$~8 a.u. and before the single positron (electron) limit is reached~\cite{AP_Arponen, PRL_Drummond}. Therefore, we have interpolated \EepTwoC{}  when one or both densities are small but finite, i.e. 8 $<$ $r_s$  $<$ 20~a.u. The asymptote of \EepTwoC{}  is $\lbrace1/(n_eV^{BN}_{ep}[n_p])+1/(n_pV^{BN}_{ep}[n_e])\rbrace^{-1}$ where \VBN{} is the parametrization given by Boronski and Nieminen~\cite{PRB_Boronski} for the single positron or electron limit. The interpolation and its first functional derivatives are continuous everywhere. Finally, we cut \EepTwoC{} when both the electron and positron densities are vanishingly small, beyond $r_s$ $>$~20 a.u.

\section{Results}~\label{sec_results}
The ionization energies of H and He are 0.5~a.u. and 0.90369~a.u., respectively. They are well above the binding energy of Ps (0.25~a.u.), therefore the electrons, as shown in figure~\ref{fig1}, remain tightly bound to the nuclei without an appreciable polarization. The positron is completely delocalized in these systems. The ionization energy, 0.34242~a.u., of the closed 2s orbital of Be is only slightly larger than the Ps binding energy so that the atom becomes polarized and the positron is bound by the induced (dynamic) dipole of Be. On the other hand, the Li ionization energy of 0.198~a.u. is lower than the binding energy of Ps and the positron forms a Ps cluster with the Li 2s electron~\cite{JPB_Mitroy}. In HPs and LiPs, the electron in Ps forms a strong chemical bond with the unpaired s electron of the atom, keeping Ps as a distinguishable unit. The formation of a Ps cluster is manifested at r $>$ 7~a.u. as the overlap of the electron and positron densities of \Lip{}, HPs and LiPs.

\subsection{Effective potentials}
We introduce now an effective single-particle potential \Veff{} using our many-body results and in Chapter~\ref{sec_discussion} we will propose it as the starting point to describe positron and Ps states in condensed matter. We invert a single-particle Schr\"odinger equation using the positron densities of the interacting systems and obtain 
\begin{equation}\label{eq3}
V_{eff}(r) = E_{eff} + \frac{1}{2M_{eff}}\frac{\nabla^2\sqrt{n_+(r)}}{\sqrt{n_+(r)}}.
\end{equation}
The effective energy \Eeff{} is the interaction energy of the asymptotic state in table~\ref{tab1}. For \Lip{}, HPs and LiPs the effective mass \Meff{} is the mass of Ps (\mPs{}) and for the other systems it is the mass of the positron (\me{}). \Veff{} is a single-particle potential for the positron even in systems where Ps is formed. We remark when the effective potential describes a system comprising a Ps atom by naming it as the Ps \Veff{}. The introduced potential is equivalent to the exact Kohn-Sham potential for a single positron with effective mass \Meff{}. \Eeff{} ensures that its asymptotic value far from the nucleus is zero also for systems including Ps. 

We also define a single-particle mass-normalized Ps effective potential $V_{eff'}^{Ps} = 2 E_{int}^{Ps} + \nabla^2(\sqrt{n_+})/(2\sqrt{n_+}$) with the effective mass \me{}. The densities obtained by solving the Schr\"odinger equation with the mass-normalized potential are the same as those of \Veff{} and the energies are multiplied by a factor of 2. In the present work, we use the mass-normalized potential to compare the Ps \Veff{} to the corresponding positron DFT potential.  

According to figure~\ref{fig3}, when r $\lesssim$ 1~a.u. the positron-nucleus Coulomb repulsion dominates over the electron-positron attractive mean-field and correlation potentials.  \Veff{} becomes attractive at larger separations, when the electron-positron correlation is comparable to the positron-nucleus Coulomb repulsion. The repulsive core of \Veff{} range from r $\lesssim$ 1.2~a.u. for \Hp{} and \Hep{} and r $\lesssim$ 2-3~a.u. for \Lip{} and \Bep{}. Although the attractive \Veff{} well is slightly deeper for \Hp{} than for \Hep{}, due to the larger polarizability of H, it is very shallow for both unbound systems. For bound \Bep{} the minimum of \Veff{} is deeper, -84.58\emthree{}~a.u. at r=3.18~a.u. The minimum value of the Ps \Veff{} potential of \Lip{} is shallower, -24.57\emthree{}~a.u. at r=4.62~a.u., but the potential well extends longer distances. Finally, the attractive Ps \Veff{} wells of the strongly bound HPs and LiPs are deep, -0.280~a.u. and -0.102~a.u., respectively. 

To show that \Veff{} can predict the correct positron density and interaction energy, we have calculated the positron (Ps) binding energy to Be (Li$^+$) by solving numerically the radial single-particle Schr\"odinger equation. For the ground state it reduces to the one-dimension problem 
\begin{equation}\label{eq2}
-\frac{1}{2M_{eff}}\frac{d^2 U}{dr^2} + V_{eff}U = EU,
\end{equation}
where $U = r\Psi$ and $\Psi$ is the s-type wavefunction. The boundary conditions for $U$ are $U(r=0)$=0 and $U(r\rightarrow\infty)$=0. For \Lip{}, HPs and LiPs the effective potentials are the Ps \Veff{} potentials and \Meff{}=2\me{}. The resulting binding energy and \rp{} given by equation~\ref{eq2} are, $E_b$ = 2.414\emthree{}~a.u. and \rp{}=10.213~a.u. for \Lip{}, $E_b$ = 2.33\emthree~a.u. and \rp{} = 11.104~a.u. for \Bep{}, $E_b$=39.210\emthree~a.u. and \rp{} = 3.673~a.u. for HPs, and $E_b$ = 10.394~a.u. and \rp{} = 6.457~a.u. for LiPs. 

\subsection{Scattering lengths}
In order to study the adequacy of \Veff{} to model positron and Ps states, we consider a positron or Ps scattering off light atoms. Many-body calculations of the s-wave phase shifts (\PhaseShift{}) and scattering lengths (\ScatLength{}) exist for \Hp{}, \Hep{} and \Bep{}. Zhang et al.~\cite{PRA_Zhang2} used the stabilized ECG-SVM to calculate the positron \ScatLength{} of H and He. Houston et al.~\cite{PRA_Houston} applied Hylleraas wavefunctions and the Kohn variational method to positrons scattering off H and Bromley et al.~\cite{JPBAMOP_Bromley} studied positron scattering off Be using polarized orbital wavefunctions. Ps scattering off  Li$^+$ ion has been studied by Mitroy and Ivanov~\cite{PRA_Mitroy2} using the stabilized ECG-SVM. 
\begin{figure*}
\begin{center}
\includegraphics[width=16cm]{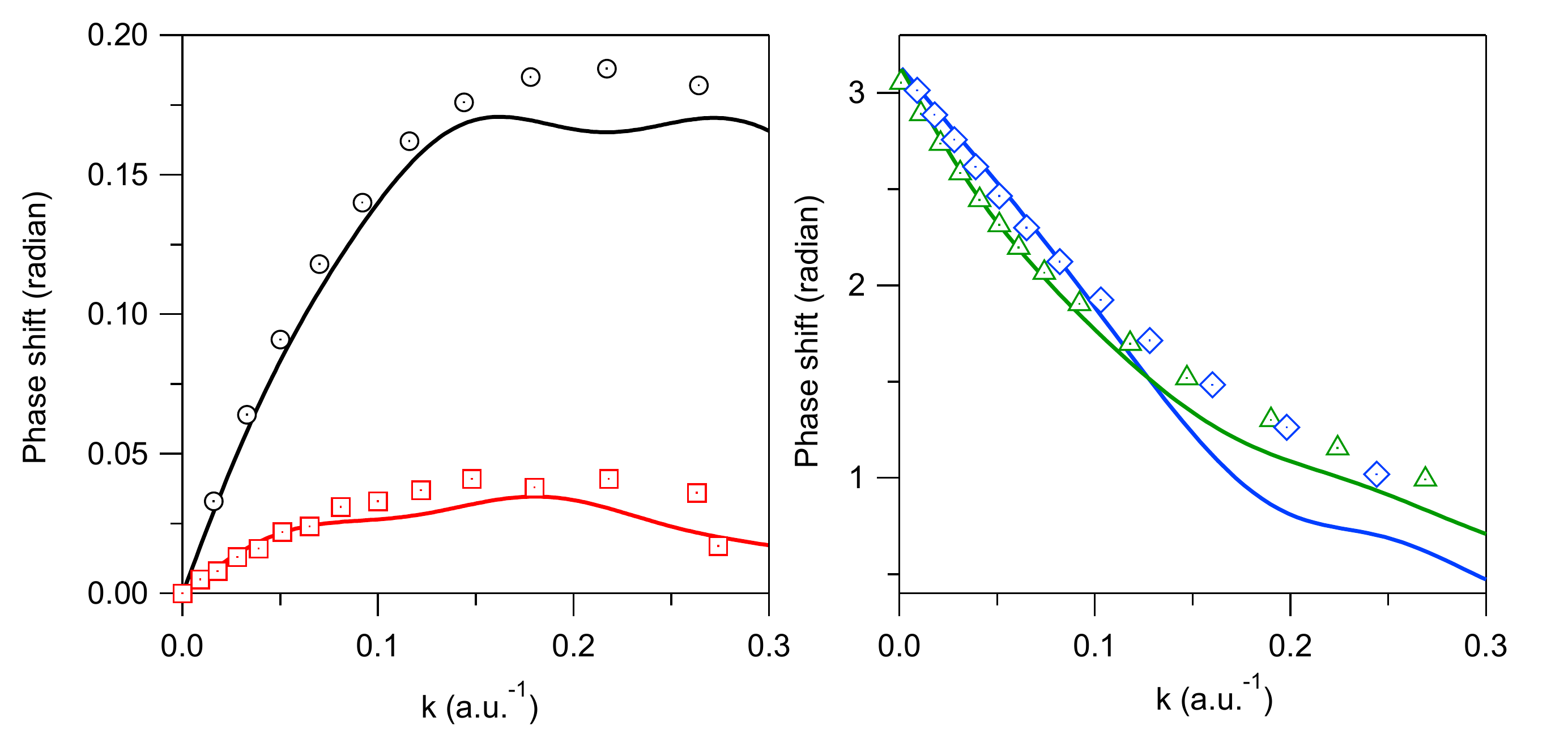}
\caption{(Color online) s-wave phase shifts for positrons scattering off H (black curve), He (red curve) and Be (green curve) and Ps scattering off Li (blue curve). The many-body values obtained by Zhang et al.~\cite{PRA_Zhang2} for H (black circles) and He (red squares), Mitroy et al.~\cite{PRA_Mitroy2} for Li (blue diamonds) and Bromley et al.~\cite{JPBAMOP_Bromley} for Be (green triangles) are also shown.} 
\label{fig9}
\end{center}
\end{figure*}

Here we calculate \PhaseShift{} and \ScatLength{} using the corresponding \Veff{} or Ps \Veff{} and compare them to the many-body values in the literature. For a positron scattering off Li, the Ps formation channel is open at all energies~\cite{PRA_Basu,JPBAMOP_McAlinden} and therefore the Ps \Veff{} \ScatLength{} and \PhaseShift{} are compared to the many-body values of Ps scattering off Li$^+$. We obtain the s-wave scattering wavefunction for a positron with the energy $E=k^2/2M_{eff}$ by solving equation~\ref{eq2}. At large distances from the nucleus the wavefunction has the form 
\begin{equation}\label{eq_scatt_wavefunc}
\lim_{r\longrightarrow\infty} \psi_0 = \frac{\sin\left(kr+\delta_0\right)}{kr}. 
\end{equation}
The wavefunction calculated numerically is fitted to this asymptote to obtain $\delta_0$ as a function of $k$. \ScatLength{} is then calculated at the low-energy limit from $k \cot\delta_0 = -1/a_0 + O(k^2)$.

The calculated \PhaseShift{}$(k)$ are plotted in figure~\ref{fig9}. They show a good agreement with the many-body values for $k$ $\lesssim$ 0.1~a.u.$^{-1}$ which suggests that (Ps) \Veff{} will remain valid to describe quasi-thermalized positrons at room temperature. For larger momenta the dynamical correlation becomes important and our values are systematically slightly lower. For Ps scattering off Li$^+$ the difference is the largest, 0.3-0.4 radians, because both the target and the projectile are deformed. For a positron scattering off Be the agreement is very good, considering that the positron binding energy to Be is 0.8\emthree{}~a.u. smaller (26\%) than the best many-body value~\cite{JAMS_Mitroy} and 0.5\emthree{}~a.u. smaller (16\%) than the binding energy by Bromley et al.~\cite{JPBAMOP_Bromley} However the \ScatLength{} value, see table~\ref{tab3}, shows the largest mismatch with the reference many-body value. For H, He and Li the present \ScatLength{} are comparable to the many-body values. 
\begin{table}[h!]
\caption{
Positron (\Hp{}, \Hep{} and \Bep{}) and Ps (\Lip{}) scattering lengths. The first column shows the values computed using \Veff{} and the last column are many-body calculations from the literature. All the values are given in a.u.
}
\label{tab3}
\begin{ruledtabular}
\begin{center}
\begin{tabular}{lcc}
\Hp{}&-1.86&-2.094~\cite{PRA_Zhang2}, -2.10278~\cite{PRA_Houston}\\
\Hep{}&-0.55&-0.474~\cite{PRA_Zhang2}\\
\Bep{}&18.76&16~\cite{JPBAMOP_Bromley}\\
Ps-Li$^+$&12.19&12.9~\cite{PRA_Mitroy2}\\ 
\end{tabular}
\end{center}
\end{ruledtabular}
\end{table} 

\subsection{Two-component DFT}~\label{sec_solid} 
2C-DFT is the basis of efficient predictive modeling of positron states in condensed matter. In this section, we study to which extent 2C-DFT within LDA is able to describe the bound states of a positron and Ps interacting with an atom. The analysis of \Lip{}, HPs, and LiPs allows us to draw conclusions also about systems including a Ps cluster. Using the vanishing positron-density limit for the electron-positron correlation energy and potential~\cite{AP_Arponen, PRL_Drummond} the LDA 2C-DFT doesn't predict the binding of positrons to atoms. We use instead \EepTwoC{}, a LDA functional that depends on the electron and positron densities and it predicts the formation of bound atom-positron states. 

\begin{figure*}
\begin{center}
\includegraphics[width=16cm]{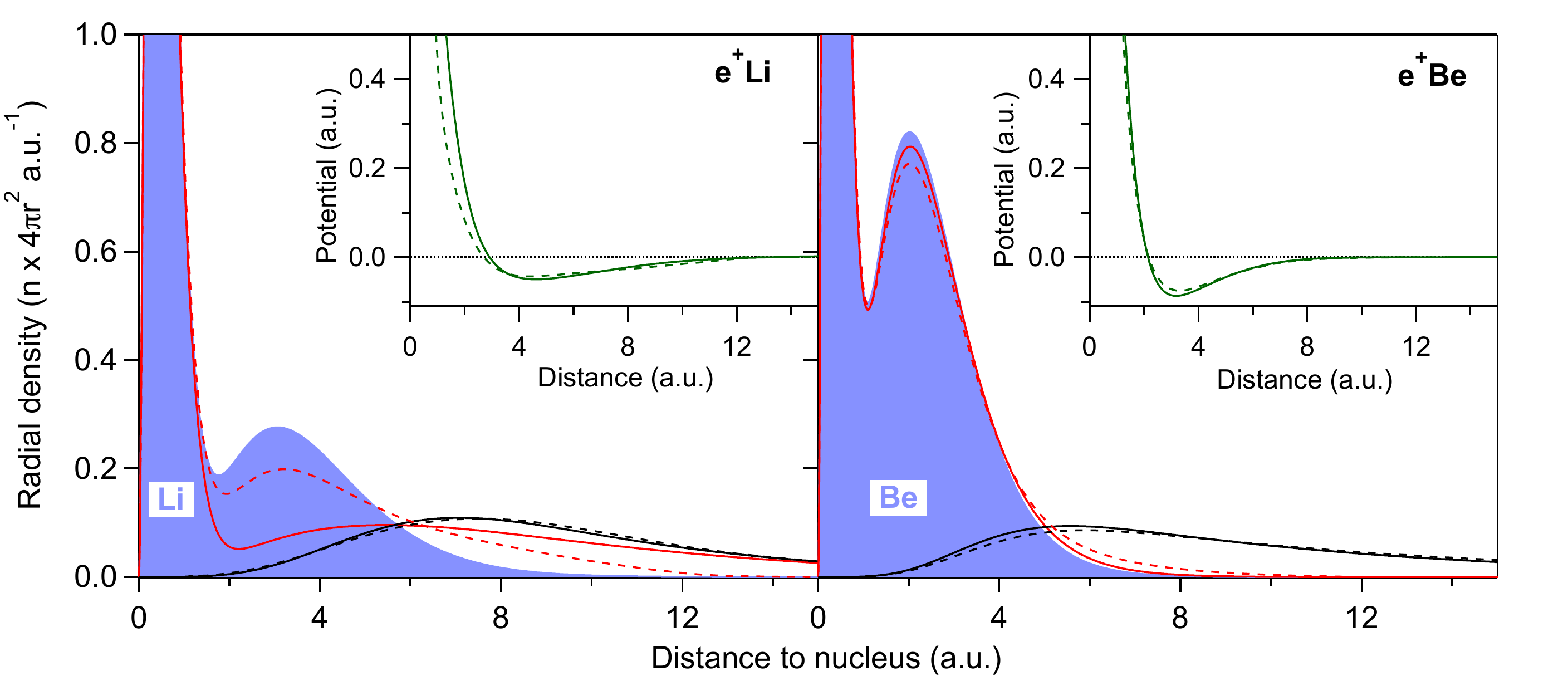}
\caption{(Color online) Many-body (full curves) and LDA (broken curves) densities and positron potentials for \Lip{} and \Bep{}. The radial electron densities of the isolated atoms (filled blue curves), and the interacting positron-atom systems (red curves), as well as the positron densities (black curves) are represented in the main panel. The insets compare the single-particle positron potentials. For \Lip{} the mass normalized potential has been plotted.} 
\label{fig8}
\end{center}
\end{figure*}

Overall, the LDA 2C-DFT predicts accurate positron densities and potentials comparable to the many-body results. Figure~\ref{fig8} compares the LDA and the many-body electron and positron densities of \Lip{} and \Bep{}. The LDA positron density of \Bep{} matches the many-body density whereas the LDA electron density is slightly more delocalized than the many-body density. The LDA positron density of \Lip{} is also accurate, however, the LDA electrons are more tightly bound to the nucleus than in the many-body calculation. The potential wells of the LDA positron potentials match \Veff{} of \Bep{} and the mass-normalized \VeffPsA{}of \Lip{}. Close to the Li nucleus the LDA positron potential is less repulsive than \Veff{} but the effect on the positron density is minor. Although \VepTwoC{} is small compared to the mean-field Coulomb potential, it is necessary to obtain a bound state for \Lip{} and \Bep{}. 

\begin{figure*}
\begin{center}
\includegraphics[width=16cm]{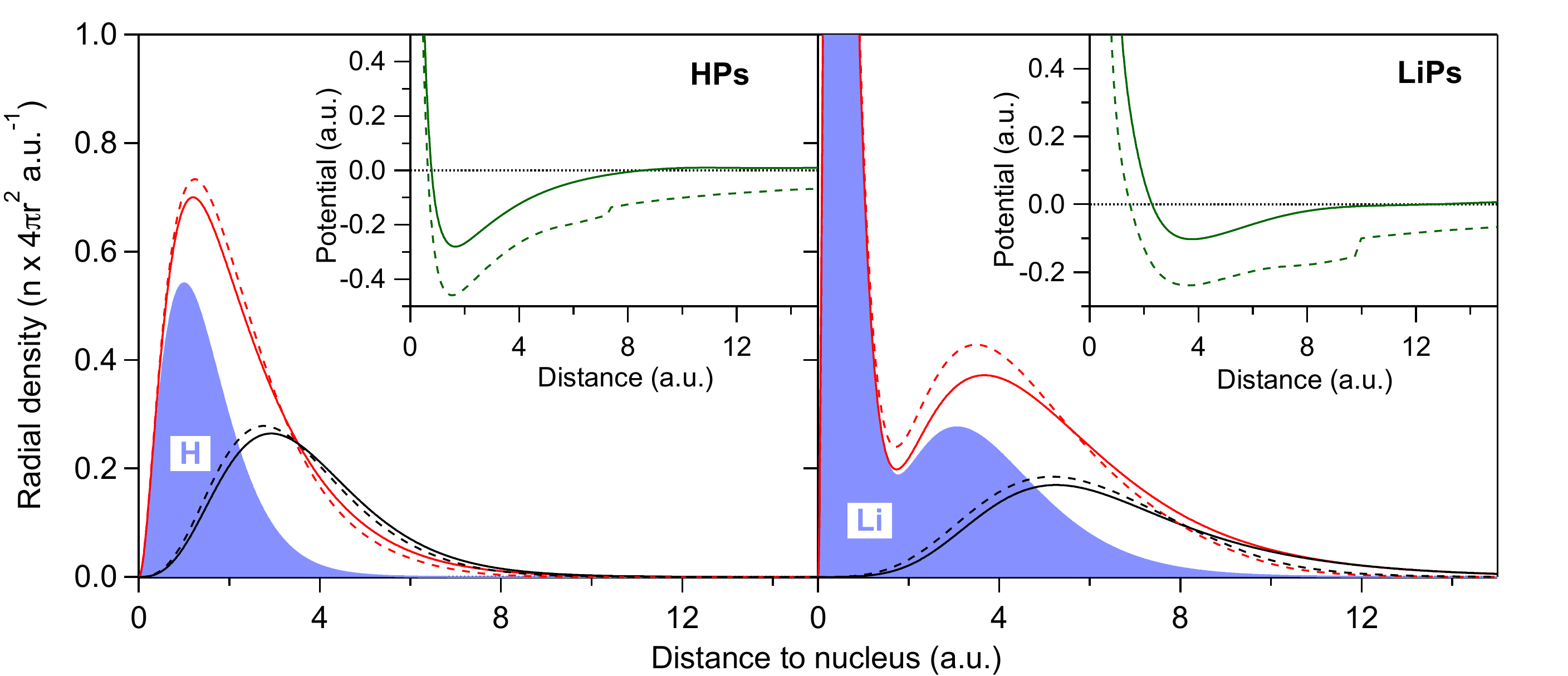}
\caption{(Color online) Many-body (full curves) and LDA (broken curves) densities and potentials for HPs and LiPs. The radial electron densities of the isolated atoms (filled blue curves), and the interacting positron-atom systems (red curves), as well as the positron densities (black curves) are represented in the main panels. The insets compare the LDA positron potential and the mass-normalized \VeffPsA{}.} 
\label{fig2}
\end{center}
\end{figure*}

In the case of HPs and LiPs the mean-field Coulomb potential alone is able to predict the formation of a bound state but including \VepTwoC{} increases the accuracy of the positron density. For both systems the LDA positron potential wells are deeper than the corresponding \Veff{} but their widths are similar up to distances, r$\sim$10~a.u. (HPs) or $\sim$12~a.u. (LiPs), where the positron densities of the bound states are already negligible. Figure~\ref{fig2} shows that the LDA electron and positron densities are slightly more localized than the many-body densities in both systems. The kinks in the LDA positron potentials of HPs and LiPs are caused by the cut-off imposed to the potential at low densities. Without the cut-off, the potentials have a long-range attractive tails which make the positron densities too delocalized. Many-body calculations at the low-density range of the electron-positron plasma would be required to obtain an electron-positron correlation potential which is accurate beyond the metallic density regime.

The asymmetric behavior of the LDA electron and positron densities with respect to the many-body calculations reflects the means of DFT to describe correlations in the interacting many-body system~\cite{PRL_Saarikoski}. The electron self-interaction causes the 2s orbital of \Lip{} to be poorly described in DFT. Moreover, 2C-DFT within LDA cannot describe accurately strongly-correlated systems like Ps. Accordingly, the LDA densities of \Lip{} don't show the formation of Ps. However, in HPs and LiPs at long separations the electron and positron densities overlap as expected when Ps forms. Overall, figures~\ref{fig8} and~\ref{fig2} show convincingly that the electron-positron correlation potential derived from the energy of an electron-positron plasma yields surprisingly accurate positron densities in bound positronic atoms, including systems where Ps forms. 

The positron binding energies of all the studied systems are only qualitative, reflecting the general inadequacy of LDA to accurately describe binding between atoms. Moreover, it is a well known problem that DFT within LDA is not able to describe dispersion interactions~\cite{JPOC_Johnson}. 

\section{\uppercase{V}$_{\textrm{eff}}$ for positron and \uppercase{P}\lowercase{s} states in condensed matter}~\label{sec_discussion}
%
It is well established that the LDA 2C-DFT yields reliable densities for positrons trapped at vacancies inside metals and semiconductors~\cite{PRB_Puska3}. To simplify the calculations or to compare different approaches, it would be desirable to calculate the positron potentials also as superpositions of atomic or molecular \Veff{} in condensed matter. However, the transferability of \Veff{} deduced from single positronic atoms or molecules is of concern. The trapping of positrons in vacancies inside metals and semiconductors occurs partly because the valence electrons relax into the vacancy as attracted by the positron increasing the binding energy and the degree of localization of the positron. In the 2C-DFT this is taken into account through the electron-positron correlation functional which lowers the energy of the positron inside the vacancy. However, in \Veff{} obtained from an atom-positron system the valence electrons remain bound to the atom and its atomic superposition cannot predict the positron trapping inside vacancies of crystalline solids. The utility of \Veff{} will not be limited by this problem in condensed matter systems where the electronic structures of the constituent atoms or molecules remain nearly undisturbed like in molecular soft condensed matter and liquids where inter-molecular interactions are weak. 

The superposition of molecular \Veff{} potentials is particularly interesting from the point of view of studying Ps embedded in molecular materials like polymers, liquids or biostructures. Typically, the exchange repulsion between the Ps and the HOMO-LUMO gap ($\sim$0.5~a.u.) of closed shell molecules prevents the formation of a Ps bound state. Instead Ps is localized in open volume pockets at the potential wells induced by the surrounding molecules. The applicability of the proposed scheme in atomic models of the material requires that \Veff{} can be derived for the molecules forming the material. The calculation of \Veff{} requires high quality many-body positron densities, which is computationally demanding with the present computing capacity. Smaller systems like HePs can be studied~\cite{PRA_Zubiaga, Unpublished_Zubiaga}, instead. He does not bind Ps due to its closed shell structure and its low polarizability. It possesses a HOMO-LUMO gap in a spin-compensated electron structure similarly to molecular matter and thus it provides a good model system to study the interaction of Ps. The knowledge gained studying model systems would allow building \Veff{} when ab-initio methods cannot be used. Moreover, our notion that the computationally efficient 2C-DFT within LDA reproduces accurately the many-body single-particles potentials for systems with Ps, raises the expectation that it could be used to construct Ps \Veff{}.

\section{Conclusions}\label{sec_conclusions} 
We have calculated the ECG-SVM many-body wavefunctions for positronic systems including a light atom (H, He, Li and Be) and a positron or Ps. Based on these results we have proposed an effective single-particle positron potential by inverting the single-particle Schr\"odinger equation arising from the many-body positron density. \Veff{} is a single-particle potential for the positron interacting with an atom which includes the full many-body correlations and it also describes a positron inside a Ps atom. The many-body positron densities and binding energies are, by construction, predicted by the introduced potential. The scattering lengths are consistent with the many-body values in the literature and the s-wave phase shifts show also good agreement for moments $k \lesssim 0.1$~a.u.$^{-1}$. The low-energy correlations are well described by \Veff{} up to energies larger than that of quasi-thermalized positrons and Ps at room temperature. The success of \Veff{} to describe the positron when a Ps complex forms, suggests that the potential can be also a valid single-particle description for the low-energy (quasi-thermalized) positron forming Ps without solving the Schr\"odinger equation for the many-body system. This possibility should be further studied in connection with Ps interacting with molecular systems. The superposition of atomic or molecular \Veff{} to calculate the positron potentials and the ensuing positron distributions in molecular condensed matter can be a valid description of the positron in Ps when the inter-molecular interactions are weak and the transferability is not of concern. 

We have also shown that the positron densities are well described within the LDA 2C-DFT for bound \Lip{}, \Bep{}, HPs and LiPs when the finite positron-density functional is used for the electron-positron correlation energy. The self-consistent LDA 2C-DFT positron potentials reproduce the binding potential well of \Veff{} accurately and predict the many-body positron densities. Although LDA 2C-DFT is less consistent predicting the electron densities, our results indicate that it yields good positron distributions also for Ps bound to atoms. This result opens the possibility to use 2C-DFT also to describe Ps interacting with extended systems. However, the need for accurate treatment of the correlations for low-density electron-positron plasmas calls for further many-body studies.

\begin{acknowledgments}
This work was supported by the Academy of Finland through the individual fellowships and the centre of excellence program. We acknowledge the computational resources provided by Aalto Science-IT project. Thanks are due to K. Varga for providing us the ECG-SVM code used in this work, to I. Makkonen for insightful discussions and to the referees for the valuable comments that improved the manuscript. 
\end{acknowledgments}


%

\end{document}